\documentclass[twocolumn,showpacs,preprintnumbers,amsmath,amssymb,nofootinbib]{revtex4}

\usepackage{graphicx}
\usepackage{color}
\usepackage{dcolumn}
\usepackage{bm}
\newtheorem{st}{S.\!\!}

\begin{document}

\preprint{APS/123-QED}

\title{Positioning with stationary emitters\\
in a two-dimensional space-time}

\author{Bartolom\'{e} Coll}
 \email{bartolome.coll@obspm.fr}
\affiliation{%
Syst\`emes de r\'ef\'erence relativistes, SYRTE-CNRS,\\
Observatoire de Paris, 75014 Paris, France.
}%

\author{Joan Josep Ferrando}
 \email{joan.ferrando@uv.es}
\author{Juan Antonio Morales}%
 \email{antonio.morales@uv.es}
\affiliation{%
Departament d'Astronomia i Astrof\'{\i}sica, \\Universitat de
Val\`encia, 46100 Burjassot, Val\`encia, Spain.
}%

\date{\today}

\begin{abstract}
The basic elements of the relativistic positioning systems in a
two-dimensional space-time have been introduced in a previous work
[Phys. Rev. D {\bf 73}, 084017 (2006)]
where geodesic positioning systems, constituted by two geodesic
emitters, have been considered in a flat space-time. Here, we want
to show in what precise senses positioning systems allow to make
{\em relativistic gravimetry}. For this purpose, we consider
stationary positioning systems, constituted by two uniformly
accelerated emitters separated by a constant distance, in two
different situations: absence of gravitational field (Minkowski
plane) and presence of a gravitational mass (Schwarzschild plane).
The physical coordinate system  constituted by the electromagnetic
signals broadcasting the proper time of the emitters are the so
called {\em emission coordinates}, and we show that, in such
emission coordinates, the trajectories of the emitters in both
situations, absence and presence of a gravitational field, are
identical. The interesting point is that, in spite of this fact,
particular additional information on the system or on the user
allows not only to distinguish both space-times, but also to
complete the dynamical description of emitters and user and even
to measure the mass of the gravitational field. The precise
information under which these dynamical and gravimetric results
may be obtained is carefully pointed out.
\end{abstract}


\maketitle

\section{Introduction}
\label{intro}

A {\em relativistic positioning system} is defined by four clocks
$\gamma_i$ ({\it emitters}) in arbitrary motion broadcasting their
proper times  $\tau^i$ in some region of a (four-dimensional)
space-time \cite{cfm-a,coll-1}. The future light cones of the points
$\gamma_i(\tau^i)$ constitute coordinate hypersurfaces
$\tau^i=constant$ of a coordinate system for that region. Indeed, at
every event of the domain, four of these cones broadcasting the
times $\tau^i$ intersect, endowing thus the event with the {\it
emission coordinates} $\{\tau^i\}:$ the four proper time signals
received by any observer at the event from the four
clocks.\footnote{The first to propose the physical construction of
emission coordinates by means of broadcast light signals seems to
have been Coll \cite{coll-2}.
 Bahder \cite{bahder}, Rovelli
\cite{Rovelli}, Blagojevi\`{c} \cite{blago} and, more recently,
Lachi\`eze-Rey \cite{rey} have considered applications of emission
coordinates to different situations. For a detailed analysis of
these references and other ones related with null coordinates see
\cite{cfm-a}.}

As a {\em location system} (i.e. as a physical realization of a
mathematical coordinate system) \cite{coll-1, coll-3}, the above
positioning system presents interesting qualities and, among them,
those of being {\em generic}, (gravity-){\em free} and {\em
immediate}. This fact has been pointed out elsewhere \cite{cfm-a,
coll-1, coll-2, coll-3} and some explicit results have been recently
obtained for the generic four-dimensional case
\cite{coll-pozo-1,coll-pozo-2,pozo}.

A full development of the theory for this generic case requires a
hard task and a previous training on simple and particular
situations. In \cite{cfm-a} we have presented a two-dimensional
approach to relativistic positioning systems introducing the basic
features that define them. We have obtained the explicit
relation between emission coordinates and any given null
coordinate system wherein the proper time trajectories of the
emitters are known. We have also developed in detail the positioning
system defined in flat space-time by geodesic
emitters. Finally, we have shown that, in arbitrary
two-dimensional space-times, the data that a user obtains from the
positioning system determine the gravitational field and its
gradient along the emitters and user world lines.

This two-dimensional approach has the advantage of allowing the use
of precise and explicit diagrams which improve the qualitative
comprehension of general four-dimensional positioning systems.
Moreover, two-dimensional scenarios admit simple and explicit
analytic results. These advantages encourage us in proposing and
solving new two-dimensional problems, many of them already suggested
in a natural way by the results presented in \cite{cfm-a}.

In the present work we want to detect the possibility of making
relativistic gravimetry or, more generally, the possibility of
obtaining the dynamics of the emitters and/or of the user, as well
as the detection of the absence or presence of a gravitational field and its
measure. This possibility is here examined by means of a (non geodesic) {\em
stationary positioning system}, that is to say a positioning
system whose emitters are uniformly accelerated and the radar
distance from each one to the other is constant. Such a stationary
positioning system is constructed in two different scenarios:
Minkowski and Schwarzschild planes.

In both scenarios, and for any user, the trajectory of the
emitters in the {\em grid}, i.e. in the plane
$\{\tau^1\}\times\{\tau^2\},$ are two parallel straight lines. At
first glance, this fact would seem to indicate the impossibility
to extract dynamical or gravimetric information from them, but
this appearance is deceptive.

We shall prove that the simple qualitative information that the
positioning system is stationary (but with no knowledge of the
acceleration and mutual radar distance of every emitter) and that
the space-time is created by a given mass (but with no knowledge
of the particular stationary trajectories followed by the
emitters) allows to know the actual accelerations of the emitters,
their mutual radar distances and the space-time metric in the
region between them in emission coordinates $\{\tau^i\}.$ Also,
the transformations to other standard coordinate
systems\footnote{For example, the coordinate transformation
relating emission coordinates to Schwarzschild coordinates, or to
inertial ones in the case of vanishing mass.} may be obtained.

The important point for gravimetry is that, in the above context,
the data of The Schwarzschild mass may be substituted by that of the
acceleration of one of the emitters. Then, besides all the above
mentioned results, including the obtaining of the acceleration of
the other emitter, the actual The Schwarzschild mass of the
corresponding space-time may be also calculated.

These relatively simple two-dimensional results strongly suggests
that relativistic positioning systems can be useful in gravimetry
at least when parameterized models for the gravitational field may
be proposed.

A comment about the presentation of the results. The user is
supposed to continuously receive data from the positioning system
and have some a priori information about it and/or the space-time.
Because of their operational importance, the statements involving
exclusively consequences of these two types of knowledge, have been
emphasized and numbered.

The work is organized as follows. In Sec. \ref{section-previous}
we summarize the basic concepts and notation about relativistic
positioning systems in a two-dimensional space-time. Sections
\ref{section-III} and \ref{section-IV} are devoted to the analysis
of stationary positioning systems respectively in Minkowski and in
Schwarzschild planes. In sections \ref{section-gravimetry} and
\ref{section-grav-along} we analyze the roles that different
user's data play in obtaining the gravitational field and in
recovering the characteristics of the positioning system. Finally,
we finish with a short discussion about the present results and
comments on prospective work in Sec. \ref{discussion}.

\section{Summary of previous results}
\label{section-previous}

In a two-dimensional space-time, let $\gamma_1$ and $\gamma_2$ be
the world lines of two clocks measuring their proper times $\tau^1$
and $\tau^2$ respectively. Suppose they broadcast them by means of
electromagnetic signals, and that the signals from each one of the
world lines reach the other. This system of two clocks ({\em
emitters}) is called a {\em relativistic positioning system}.

A domain $\Omega$ in the region between both emitters constitutes
a {\em coordinate domain}. Indeed, the past light cone of every
event in $\Omega$ cuts the emitter world lines at
$\gamma_1(\tau^1)$ and $\gamma_2(\tau^2)$, respectively. Then
$(\tau^1,\tau^2)$ are the coordinates of the event: the two proper
time signals received by any observer at the event from the two
clocks. We shall refer to them as the {\em emission coordinates}
$\{\tau^1,\tau^2\}$ of the system in $\Omega$. Also, by
continuity, we shall call {\em emission coordinates of the
emitters} the pair of proper times formed by the proper time of
the clock in question and that received from the other clock.
Nevertheless,  the signals $\tau^1$ and $\tau^2$ do not constitute
coordinates for the events in the outside region.

The plane $\{\tau^1\}\times\{\tau^2\}$ ($\tau^1, \tau^2 \in
\mathbb{R}$) in which the different data of the positioning system
can be transcribed is called, as already remembered, the  grid of
the positioning system. In this grid, the trajectories of the two
emitters (which, for the auto-locating positioning systems
considered below, can be drawn from the positioning data broadcast
by the emitters) define an interior region and two exterior ones.
This interior region in the grid is in one-to-one correspondence
with the interior region in the space-time, i.e. with the set of
events that can be distinguished by the pair of times
$(\tau^1,\tau^2)$ that reach them. But  the exterior regions in the
grid are not in one-to-one correspondence with the corresponding
exterior regions in the space-time, because these last regions are
foliated by null geodesics whose events have all the same values
$(\tau^1,\tau^2)$ that the emission coordinates of the emitter that
borders the region. So, the pairs $(\tau^1,\tau^2)$ of the exterior
regions in the grid {\em are not} emission coordinates: they do not
correspond to pairs of signals broadcast by the two emitters.
Consequently, for  this emission protocol, they do not correspond to
points of the two-dimensional space-time (nevertheless, as
mathematical extensions, they could be of interest in some
situations). The grid origin depends exclusively on the choice of
the instant zero of the emitter clocks. This synchronization is not
a necessary prior information for positioning and it can be obtained
from the system data.

Emission coordinates are null coordinates and thus the
space-time metric depends on the sole {\em metric function} $m$:
\begin{equation} \label{metric-taus}
ds^2 = m(\tau^1,\tau^2) d \tau^1 d\tau^2
\end{equation}

An observer $\gamma,$ traveling throughout an emission coordinate
domain $\Omega$ and equipped with a receiver allowing the reading of
the received proper times $(\tau^1,\tau^2)$ at each point of his
trajectory, is called a {\it user} of the positioning system.

From now on we will consider {\em auto-locating positioning
systems}, which are systems in which every emitter clock not only
broadcasts its proper time but also the proper time that it receives
from the other, i.e. the systems that broadcast the emission
coordinates of the emitters. Thus, the physical components of an
auto-locating positioning system are \cite{cfm-a} a {\em spatial
segment} constituted by two emitters $\gamma_1$, $\gamma_2$
broadcasting their proper times $\tau^1,$ $\tau^2$ {\em and} the
proper times $\bar{\tau}^2$, $\bar{\tau}^1$ that they receive each
one from the other, and a {\em user segment} constituted by the set
of all users traveling in an internal domain $\Omega$ and receiving
these four broadcast times $\{\tau^1, \tau^2; \bar{\tau}^1,
\bar{\tau}^2\}$.\footnote{Observe that, unlike classical positioning
systems, such as the Global Positioning System, it does not exist
here a `control segment', because relativistic emission coordinates
constitute epistemologically a {\em primary} coordinate system, in
contrast with classical positioning, which uses the emitters simply
as moving beacons to transmit {\em another} Newtonian reference
system, considered as principal,  like the World Geodetic System or
the International Terrestrial Reference System. Of course, a
`maintenance segment' keeping the emitters in working order would be
necessary for technical but not scientific, reasons. See
\cite{coll-3} for some details. It remains always possible to
couple, for complementary practical purposes, a reference system to
a relativistic positioning one, but our purpose is to reveal the
{\em sufficiency} of  relativistic positioning systems, as defined
here, as physical realizations of coordinate systems.}

It is worth remarking that any user receiving continuously the
emitted times $\{\tau^1, \tau^2\}$ knows his trajectory in the grid.
Indeed, from these {\em user positioning data} $\{\tau^1, \tau^2\}$
the equation $F$ of the user trajectory may be extracted:
\begin{equation}  \label{user-trajectory}
\tau^2 = F(\tau^1)
\end{equation}

On the other hand, any user receiving continuously the {\em
emitter positioning data} $\{\tau^1, \tau^2; \bar{\tau}^1,
\bar{\tau}^2\}$ may extract from them not only the equation
(\ref{user-trajectory}) of  his
trajectory, but also the equations of the trajectories
of the emitters in the grid:
\begin{equation} \label{emitter-trajectories}
\varphi_1(\tau^1) = \bar{\tau}^2 \, , \qquad  \varphi_2(\tau^2) =
\bar{\tau}^1
\end{equation}

Eventually, the positioning system can be endowed with
complementary devices.  Thus, in order to obtain the dynamic
properties of the system, the emitters $\gamma_1$, $\gamma_2$
could carry accelerometers and broadcast their acceleration
$\alpha_1$, $\alpha_2$, meanwhile the users $\gamma$ could be
endowed with receivers able to read, in addition to the emitter
positioning data, also the broadcast emitter accelerations
$\{\alpha_1, \alpha_2\}$. These new elements allow any user to
know the acceleration scalar of the emitters:
\begin{equation} \label{emitter-accelerations}
\alpha_1 = \alpha_1(\tau^1) \, , \qquad \alpha_2 = \alpha_2(\tau^2)
\end{equation}

In some cases, it can be useful that the users generate their own
data, carrying a clock to measure their proper time and/or an
accelerometer to measure their proper acceleration $\alpha$. The
user's clock allows any user to know his proper time function
$\tau(\tau^1)$ (or $\tau(\tau^2)$) and, consequently by using
(\ref{user-trajectory}), to obtain the proper time parametrization
of his trajectory:
\begin{equation} \label{user-proper-time}
\gamma \equiv \begin{cases} \ \tau^1 = \tau^1(\tau) \\ \ \tau^2 =
\tau^2(\tau) \end{cases}
\end{equation}
The user's accelerometer allows any user to know his proper
acceleration scalar:
\[
\alpha = \alpha(\tau)
\]

Thus, a relativistic positioning system may generate the {\it user
data}:
\begin{equation} \label{user-data}
\{\tau^1, \tau^2; \bar{\tau}^1, \bar{\tau}^2; \alpha_1, \alpha_2;
\tau, \alpha\}
\end{equation}

The emitter trajectories (\ref{emitter-trajectories}) and the
emitter accelerations (\ref{emitter-accelerations}) do not depend
on the user that receives them. Thus, among the user data
(\ref{user-data}) we can distinguish besides the emitter
positioning data $\{\tau^1, \tau^2; \bar{\tau}^1, \bar{\tau}^2\}$,
the {\em emitter dynamical data} $\{\alpha_1, \alpha_2\}$, their
union or {\it public data} $\{\tau^1, \tau^2; \bar{\tau}^1,
\bar{\tau}^2; \alpha_1, \alpha_2\}$ and the {\it user proper data}
$\{\tau, \alpha\}$.

The purpose of the (relativistic) theory of positioning systems is
to develop the  techniques necessary to determine the space-time
metric as well as the dynamics of emitters and users by means of
physical information carried by light signals. But in order to
develop it in known space-times, it is useful to connect our
emission coordinates to the usual coordinates in which these
space-times have been already studied. We finish this section by
summarizing the explicit obtaining of the emission coordinates from
an arbitrary null coordinate system $\{\texttt{u},\texttt{v}\}$ in a
generic two-dimensional space-time \cite{cfm-a}. Let us assume the
{\it proper time history of two emitters} to be known in this
coordinate system:
\begin{equation}  \label{principalemit}
\gamma_1 \equiv \begin{cases} \texttt{u} = u_1(\tau^1) \\
\texttt{v} = v_1(\tau^1)\end{cases} \qquad \gamma_2 \equiv
\begin{cases} \texttt{u} = u_2(\tau^2) \\ \texttt{v} = v_2(\tau^2)
\end{cases}
\end{equation}

We can introduce the proper times as coordinates $\{\tau^1,\tau^2\}$
by defining the transformation to the null system
$\{\texttt{u},\texttt{v}\}$ given by:
\begin{equation}  \label{coordinatechange0}
\begin{array}{l}
\texttt{u} = u_1(\tau^1) \\
\texttt{v} = v_2(\tau^2)
\end{array}
\qquad \quad
\begin{array}{l}
\tau^1 = u_1^{-1}(\texttt{u}) = \tau^1(\texttt{u}) \\
\tau^2 = v_2^{-1}(\texttt{v}) = \tau^2(\texttt{v})
\end{array}
\end{equation}

Thus, relations (\ref{coordinatechange0}) define {\it emission
coordinates} in the {\it emission coordinate domains} $\Omega$
between both emitters. Note that outside this region of domains the
transformation (\ref{coordinatechange0}) also determines null
coordinates, but they are not emission coordinates, i.e. they cannot
be constructed by means of broadcast signals  \cite{cfm-a}.

In emission coordinates, the emitter trajectories take the
expression:
\begin{equation}  \label{principalemitintaus0}
\gamma_1 \equiv \begin{cases} \tau^1 = \tau^1 \\ \tau^2 =
\varphi_1(\tau^1)\end{cases} \qquad \gamma_2 \equiv
\begin{cases} \tau^1 = \varphi_2(\tau^2) \\ \tau^2 = \tau^2
\end{cases}
\end{equation}
where, from (\ref{principalemit}) and (\ref{coordinatechange0}), the
emitter trajectories $\varphi_i$ are given by:
\begin{equation}  \label{phis}
\varphi_1 = v_2^{-1} \circ v_1    \, , \qquad   \varphi_2 = u_1^{-1}
\circ u_2
\end{equation}

\section{Stationary positioning in Minkowski plane}
\label{section-III}

We consider here the positioning system defined by two non inertial
stationary emitters $\gamma_1$, $\gamma_2$ in absence of
gravitation, that is to say, in Minkowski plane (two-dimensional
flat space-time). Thus, the emitters have a uniformly accelerated
motion and they are at rest with respect to each other, i.e. the
hyperbolic emitter trajectories have the same asymptotes (these
observers have been largely studied by Rindler in \cite{rindler}).
Then we can choose inertial null coordinates
$\{\texttt{u},\texttt{v}\}$ such that the trajectories of the
emitters are [see Fig. \ref{fig:accelerated-1}(a)]:
\[
\texttt{v} \, \texttt{u} = - \frac{1}{\alpha_i^2}
\]
where $\alpha_i$, $0 < \alpha_1 < \alpha_2$, is the acceleration
parameter of the emitter $\gamma_i$.

It is known that the space-like half-straight lines cutting at the
coordinate origin determine the locus of simultaneous events for
both emitters. Thus, we can define a synchronization of the emitter
clocks by giving their proper times $\tau^1_0$ and $\tau^2_0$ at two
simultaneous events. With this synchronization, the proper time
history of the emitters is [see Fig. \ref{fig:accelerated-1}(a)]:
\begin{equation}  \label{III-emitters}
\! \gamma_i \equiv \begin{cases} \,\displaystyle \texttt{u} =
u_i(\tau^i) = \frac{1}{\alpha_i} \exp[\alpha_i (\tau^i- \tau^i_0)]
\\[3mm] \,\displaystyle
\texttt{v} = v_i(\tau^i) = -\frac{1}{\alpha_i} \exp[-\alpha_i
(\tau^i- \tau^i_0)]
\end{cases}
\end{equation}
where the repeated index does not indicate summation.

\subsection{Emission coordinates and emitter trajectories}

According to (\ref{coordinatechange0}), the emission coordinates
$\{\tau^1,\tau^2\}$ are defined by the transformation to the
inertial system $\{\texttt{u},\texttt{v}\}$:
\begin{equation}  \label{III-change-a}
\begin{array}{l}
\! \! \! \! \displaystyle{ \texttt{u}  = u_1(\tau^1) =
\frac{1}{\alpha_1} \exp[\alpha_1 (\tau^1 - \tau^1_0)}]
\\[3mm]
\! \! \! \! \displaystyle \texttt{v}  = v_2(\tau^2) =
-\frac{1}{\alpha_2} \exp[-\alpha_2 (\tau^2 - \tau^2_0)]
\end{array}
\end{equation}
and the inverse transformation is:
\begin{equation}  \label{III-change-b}
\begin{array}{l}
\displaystyle \tau^1 =  \tau^1_0 + \frac{1}{\alpha_1}
\ln (\alpha_1 \texttt{u}) \\[3mm]
\displaystyle \tau^2 =  \tau^2_0 -\frac{1}{\alpha_2} \ln (-\alpha_2
\texttt{v})
\end{array}
\end{equation}

From (\ref{principalemitintaus0}) and (\ref{phis}), we can obtain
the expression of the trajectories in emission coordinates:

\begin{st} \label{III-recti-trayec}
In Minkowski plane, the trajectories $\gamma_1$, $\gamma_2$ of the
emitters of a stationary positioning system in emission
coordinates $\{\tau^1,\tau^2\}$ are {\rm parallel straight lines}
of the form:
\begin{eqnarray}  \label{III-principaltaus1}
\gamma_1 & \equiv & \begin{cases} \tau^1 = \tau^1 \\ \tau^2 =
 \frac{1}{\omega}(\tau^1 - q - \sigma) \equiv \varphi_1(\tau^1)
\end{cases}
\\
\gamma_2 & \equiv & \begin{cases} \tau^1 = \omega \tau^2 - q +
\sigma \equiv \varphi_2(\tau^2)
 \\
\tau^2 = \tau^2
\end{cases}  \label{III-principaltaus2}
\end{eqnarray}
The {\em slope parameter} $\omega$, the {\em separation parameter}
$q$ and the {\em synchronization parameter} $\sigma$ are related to
the acceleration parameters $\alpha_i$ and the synchronization
instants $\tau^i_0$ of the emitters by:
\begin{eqnarray} \label{III-parametres-a}
\omega \equiv \frac{\alpha_2}{\alpha_1} > 1 , &\ & \quad  q \equiv
\frac{1}{\alpha_1} \ln \frac{\alpha_2}{\alpha_1}> 0
\\[2mm]
\sigma&\equiv&\tau^1_0 - \frac{\alpha_2}{\alpha_1} \, \tau^2_0
\quad \label{III-parametres-b}
\end{eqnarray}
\end{st}

\begin{figure}[htb]
    \includegraphics[angle=0,width=0.48\textwidth]{%
        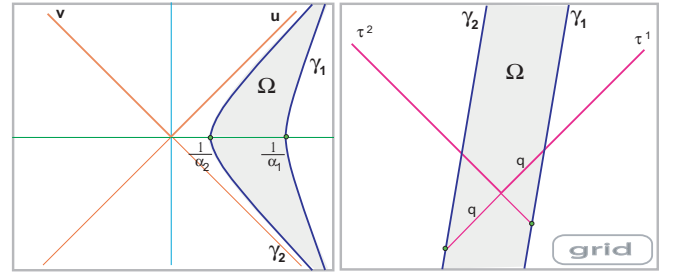}
    \caption{       \label{fig:accelerated-1}
        Trajectories of the uniformly accelerated emitters in flat
        space-time. (a) In inertial null coordinates
        $\{\texttt{u}, \texttt{v}\}$. (b) In the grid $\{\tau^1,\tau^2\}$;
        in this picture we have chosen the synchronization parameter $\sigma = 0$.}
\end{figure}

Fig. \ref{fig:accelerated-1}(b) illustrates this situation for a
vanishing synchronization parameter.

\subsection{Emitter's positioning and dynamical parameters}

Let us consider a user traveling on the emission domain $\Omega$
defined by this positioning system and receiving the emitter
positioning data $\{\tau^1, \tau^2; \bar{\tau}^1, \bar{\tau}^2\}$.
Firstly, these data, under the form of the two sequences
$\{\tau^1,\bar{\tau}^2\}$ and $\{\bar{\tau}^1,\tau^2\},$ determine
the emitter trajectories $\varphi_i(\tau^i)$ on the grid which,
according to statement \ref{III-recti-trayec}, will be straight
lines, from which the user can extract the slope, separation and
synchronization parameters $\omega > 1$, $q > 0$ and $\sigma$ (see
(\ref{III-principaltaus1}) and (\ref{III-principaltaus2})).

In fact, these parameters can be extracted from the emitter
positioning data at two different events:

\begin{st} \label{III-two-events}
In terms of the emitter positioning data $\{\tau^1_P,
\tau^2_P;\bar{\tau}^1_P, \bar{\tau}^2_P\}$ and $\{\tau^1_Q,
\tau^2_Q;\bar{\tau}^1_Q, \bar{\tau}^2_Q\}$ at two different events
$P$ and $Q$ of any user, the slope, separation and synchronization
parameter $\omega > 1$, $q > 0$ and $\sigma,$ characterizing the
parallel trajectories of the emitters in the grid
$\{\tau^1\}\times\{\tau^2\}$, are given by:
\begin{equation} \label{III-par-taus}
\begin{array}{l}
\displaystyle \omega = \frac{\Delta\bar{\tau}^1}{\Delta\tau^2}
\equiv
\frac{\bar{\tau}^1_P - \bar{\tau}^1_Q}{\tau^2_P - \tau^2_Q} \\[3.5mm]
\displaystyle q = \frac12 [\tau^1_P + \bar{\tau}^1_P +
\frac{\Delta\bar{\tau}^1}{\Delta\tau^2}(\tau^2_P +
\bar{\tau}^2_P)] \\[3.5mm]
\displaystyle \sigma = \frac12 [\tau^1_P + \bar{\tau}^1_P -
\frac{\Delta\bar{\tau}^1}{\Delta\tau^2}(\tau^2_P +
\bar{\tau}^2_P)]
\end{array}
\end{equation}
\end{st}

Once these parameters are evaluated, expression
(\ref{III-parametres-a}) implies that $\omega$ and $q$ determine
one-to-one the acceleration of the emitters and
(\ref{III-parametres-b}) shows that the parameter $\sigma$ is
necessary in obtaining the emitter synchronization:

\begin{st} \label{III-alfas-taus}
For stationary positioning systems in Minkowski plane, the
accelerations $\alpha_1$, $\alpha_2$ of the emitters may be obtained
in terms of the slope data parameter $\omega > 1$ and the separation
data parameter $q > 0,$ as:
\begin{equation} \label{III-acceleracions}
\alpha_1 = \frac{1}{q} \ln \omega \, , \qquad \alpha_2 =
\frac{\omega}{q} \ln \omega \ ,
\end{equation}
and their synchronization times $\tau^1_0$ and $\tau^2_0$ indicated
by the emitter clocks at two simultaneous events, are related to
$\omega$ and the synchronization data parameter $\sigma$ by
\begin{equation} \label{III-sigma}
\tau^1_0 = \sigma + \omega \tau^2_0
\end{equation}
\end{st}

This statement gives the parameters $\alpha_1$ and $\alpha_2$, as
well as the relation between $\tau^1_0$ and $\tau^2_0,$ in terms of
the data parameters $\omega,$ $q$ and $\sigma$, which can be
extracted from the data received by the user. Apart from their
specific physical meaning, as accelerations and synchronization
times of the emitters, these parameters $\alpha_1,$ $\alpha_2,$
$\tau^1_0$ and $\tau^2_0,$ yield an operational definition of the
null canonical inertial coordinates $\{\texttt{u},\texttt{v}\}$
associated with the emitters, i.e. those with origin in the
intersection of the null asymptotes of the stationary system to
which the emitters belong. This operational definition is offered by
relations (\ref{III-change-a}). Every set
$\{\texttt{u},\texttt{v}\}$  of null coordinates so determined by
(\ref{III-change-a}) is one-to-one related to a standard inertial
coordinate system  $\{t,x\},$  by  $\texttt{u} = t+x,$ $\texttt{v} =
t - x $, and inertial systems of fixed origin are related by a
one-parametric Lorentz transformation.  It is such a parameter (say
$\tau_0^2$ here) that relations (\ref{III-acceleracions}) and
(\ref{III-sigma}) leave free for given data parameters $\omega$, $q$
and $\sigma.$

\subsection{Metric in emission coordinates}

Now, let us analyze the metric information that any user can obtain
from the emitter positioning data. The expression of the metric
function in emission coordinates follows from the coordinate
transformation (\ref{III-change-a}):
\begin{equation}  \label{III-metric-0}
m(\tau^1, \tau^2) = u_1'(\tau^1) v_2'(\tau^2) = \exp[\alpha_1 \tau^1
- \alpha_2 \tau^2 - \alpha_1 \sigma]
\end{equation}
Consequently, by using (\ref{III-acceleracions}), we can write the
metric function in terms of the data $\{\tau^1, \tau^2\}$ and the
parameters $\{\omega, q, \sigma \}$:

\begin{st} \label{III-prop-metric}
In Minkowski plane, the space-time metric in emission coordinates of
a stationary positioning system is given, in terms of the emitter
positioning data $\{\tau^1, \tau^2; \bar{\tau}^1, \bar{\tau}^2\},$
by:
\begin{equation}
\label{III-metric} \displaystyle ds^2 = \exp\left[\frac{\ln
\omega}{q} (\tau^1 - \omega \tau^2 - \sigma)\right]  d \tau^1 d
\tau^2
\end{equation}
where the data parameters $\omega$, $q$, $\sigma$ can be obtained by
{\rm (\ref{III-par-taus})} in terms of the values of these data at
two events.
\end{st}

This statement shows that the sole public quantities $\{\tau^1,
\tau^2; \bar{\tau}^1, \bar{\tau}^2\}$ received by any user allow him
to know the space-time metric everywhere. Nevertheless, it is worth
remarking that, in order for a user to obtain this result, the user
must be informed that {\em the emitter positioning data come from a
stationary positioning system in absence of gravitational fields}.
Otherwise, the above deduction based on the existence of the
coordinate transformation (\ref{III-change-a}) would not be valid.
Remark also that this required information does not demand the
specific value of the accelerations of the emitters, which may be
also determined by the user by relations (\ref{III-acceleracions}).

The users may know that the system of the two emitter clocks is
stationary by an {\em a  priori information}, forming part of the
dynamical characteristics of the positioning system and its foreseen
control. But they may also obtain this {\em information in real
time}, if clocks and users are endowed with devices allowing the
users to know the emitter accelerations $\{\alpha_1, \alpha_2\}$ at
every instant (see comments in sections \ref{section-gravimetry} and
\ref{discussion} and Ref. \cite{cfm-c})

In any case, the user information of the dynamics of the pair of
clocks by any of these two methods is generically necessary, because
{\em emitter trajectories that are parallel straight lines in the
grid are not necessarily uniformly accelerated trajectories in a
flat space-time} \cite{cfm-c}. In Sec. \ref{section-IV} we show that
in a non flat space-time we can also find in the grid the same
picture that Fig. \ref{fig:accelerated-1}(b).

\subsection{Stationary user}

The considerations that follow in this section are qualitatively
independent of the synchronization of the emitters. For this
reason, we assume for the sake of simplicity the synchronization
parameter to be zero, $\sigma = 0$.

All the above considerations are valid for any user $\gamma$
which, as we have already said, can also obtain from the user
positioning data $\{\tau^1, \tau^2\}$ his own trajectory in the
grid. Then, the expression (\ref{III-metric}) of the metric
tensor allows him to calculate as usual his proper time lapse and acceleration in
terms of the emitter positioning data $\{\tau^1, \tau^2;
\bar{\tau}^1, \bar{\tau}^2\}$.

\begin{figure}[htb]
    \includegraphics[angle=0,width=0.48\textwidth]{%
         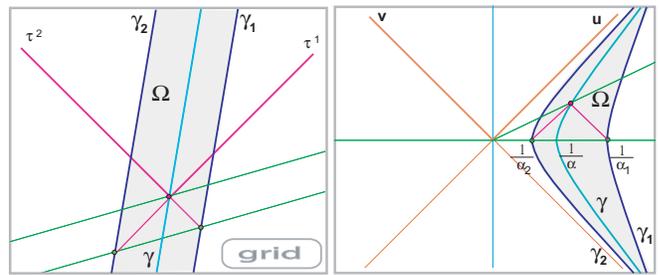}
    \caption{       \label{fig:accelerated-2}
        (a) The user positioning data $\{\tau^1,\tau^2\}$ determine the
        trajectory of the user $\gamma$ in the grid: $\tau^2 = \frac{1}{\omega}
        \tau^1$. The parallel straight lines $\tau^2 = -\frac{1}{\omega}
        \tau^1+ C$ are the locus of  simultaneous events for the uniformly
        accelerated emitters. (b) Trajectory of the user $\gamma$ in inertial null
        coordinates $\{\texttt{u}, \texttt{v}\}$: he also has an hyperbolic motion.}
\end{figure}

Suppose now that a user obtains from these data the
following trajectory:
\begin{equation} \label{III-user1-a}
\tau^2 = F(\tau^1) = \frac{1}{\omega} (\tau^1 -c) \, , \qquad -q < c
< q \, ,
\end{equation}
that is, a straight line parallel to the emitter trajectories [see
Fig. \ref{fig:accelerated-2}(a), where we plot the case $c=0$].
What is his dynamics and his relation to the emitters? The answer
is given by [see Fig. \ref{fig:accelerated-2}(b)]:

\begin{st} \label{III-prop-user-estat-acc}
 The users which find that their
trajectory in the grid is a straight line parallel to the
emitter ones {\rm [Eq. (\ref{III-user1-a})]}, are the users
which are stationary with respect to the emitters. Their
acceleration is the weighted geometric mean of the emitters'
accelerations:
\begin{equation}
\label{III-user1-b}
 \alpha \equiv \sqrt{\alpha_1^{1+c/q}\, \alpha_2^{1- c/q}},
\end{equation}
and their proper time lapse $\Delta\tau$ is, for any of the two
emitter clocks (no summation)
\begin{equation} \label{III-user-tau}
\Delta \tau =\frac{\alpha_i}{\alpha} \Delta \tau^i .
\end{equation}
\end{st}

To obtain these results, let us construct as above indicated the
canonical inertial coordinate systems $\{\texttt{u},\texttt{v}\}$
associated with the emitters. The transformation equations from the
emission coordinates $\{\tau^1, \tau^2\}$  to these inertial
coordinates are given by (\ref{III-change-a}). In them, using the
inverse transformation (\ref{III-change-b}), the trajectory equation
(\ref{III-user1-a}) becomes:
\begin{equation}  \label{III-user1-bb}
\texttt{u} \texttt{v} = - \frac{1}{\alpha^2}\, ,
\end{equation}

where the constant $\alpha$ is given by (\ref{III-user1-b}). This
means that this trajectory belongs to the stationary family of
trajectories that contains that of the two emitters, and that its
constant acceleration is $\alpha.$ To finish the proof of the
statement \ref{III-prop-user-estat-acc} we consider the proper time
parametrization of the stationary user (\ref{III-user1-bb}):
\begin{equation}  \label{III-user1-bbb}
\! \gamma \equiv \begin{cases} \,\displaystyle \texttt{u} =
u(\tau) = \frac{1}{\alpha} \exp[\alpha (\tau- \tau_0)]
\\[3mm] \,\displaystyle
\texttt{v} = v(\tau) = -\frac{1}{\alpha} \exp[-\alpha (\tau-
\tau_0)]
\end{cases}
\end{equation}
Then, (\ref{III-user1-bbb}) and the coordinate transformation
(\ref{III-change-a}) give the expression (\ref{III-user-tau}) of the
proper time lapse of the user.

\subsection{Simultaneous events for the emitters}

In order to better understand the grid associated with the emission
coordinates of our two-dimensional stationary positioning systems,
let us consider the locus of simultaneous events for both emitters,
i.e. the orthogonal lines to the stationary family of trajectories
that contains that of the two emitters which, as it is well known,
are the space-like geodesics cutting at the common point of the two
asymptotes of the emitters. How can they be constructed in the grid?
The answer is:
\begin{st} \label{III-prop-simultaneous}
The locus of simultaneous events for the emitters of a stationary
positioning system in Minkowski plane are parallel straight lines of
slope $-\frac{1}{\omega}$, the same, up to sign, as that of the
trajectory of the emitters.
\end{st}

To see this, take into account that in inertial null coordinates
these lines are $\texttt{u} = - \kappa \texttt{v} \ , \kappa > 0$.
It is then sufficient to change to emission coordinates using
(\ref{III-change-a}) to obtain:
\[
 \tau^2 = - \frac{1}{\omega} \tau^1 + C , \qquad C \equiv
\frac{1}{\alpha_2} \ln \frac{\kappa}{\omega} + 2 \tau_0^2
\]
which shows directly the statement [see Fig.
\ref{fig:accelerated-2}(a)].

\subsection{Geodesic user}

To finish this section, let us consider the user positioning data
received by an inertial (geodesic) user. For simplicity, but with no
qualitative differences, we shall choose him to be at rest with
respect to the emitters at the initial instant at which $\tau^1_0 =
\tau^2_0 = 0$ (remember that we have taken $\sigma = 0$). For him we
have:

\begin{st} \label{III-prop-inertial}
The trajectory in emission coordinates of inertial users initially
at rest with respect to  two synchronized stationary emitters is:
\begin{equation}  \label{III-user-geo}
\tau^2 = F(\tau^1) = -\frac{1}{\alpha_2} \ln[A- \omega \exp[\alpha_1
\tau^1]] \, ,
\end{equation}
where $A$ is a constant.
\end{st}

\begin{figure}[thb]
    \includegraphics[angle=0,width=0.48\textwidth]{%
        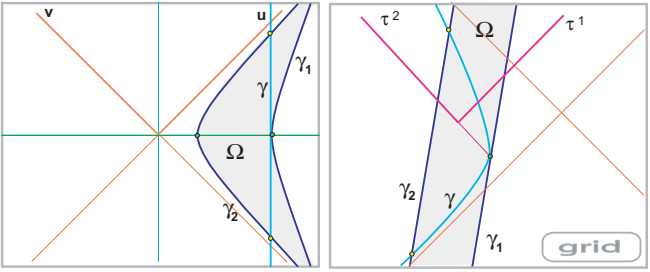}
    \caption{       \label{fig:accelerated-3}
        Trajectory of a geodesic user $\gamma$ with $x_0 = \frac{1}{\alpha_1}$:
        (a) In inertial null coordinates
        $\{\texttt{u}, \texttt{v}\}$. (b) In the grid $\{\tau^1,\tau^2\}$
        defined in flat space-time by two uniformly accelerated
        emitters.}
\end{figure}

This equation may be obtained as follows. Choose the null inertial
system $\{\texttt{u},\texttt{v}\}$ associated with the inertial
observer having the same initial instant as the two emitters, that
is to say, such that $\tau^1_0 = \tau^2_0 = 0.$ Because the user is
at rest with respect this null inertial system, its trajectory is
given by [see Fig. \ref{fig:accelerated-3}(a)]:
\begin{equation}
\texttt{v} = \texttt{u} - 2 x_0 \, , \qquad x_0 < \frac{1}{\alpha_2}
\end{equation}

Then, making use of the coordinate transformation
(\ref{III-change-a}), a straightforward calculation leads to
expression (\ref{III-user-geo}), with $A \equiv 2 x_0 \alpha_2$.
Note that the coordinate $\tau^2$ run in $] -\frac{1}{\alpha_2} \ln
A, + \infty[$ when $\tau^1$ run in $] - \infty, \frac{1}{\alpha_1}
\ln( A/\omega)\ [$. Thus the data $\{\tau^1, \tau^2\}$ of a geodesic
user lead to a trajectory in the grid with a double asymptotic
behavior [see Fig. \ref{fig:accelerated-3}(b)].

\section{Positioning in Schwarzschild plane}
\label{section-IV}

In this section we study stationary positioning systems defined  by
two emitters at rest with respect to the gravitational field in {\it
Schwarzschild plane}.

The usual expression for the space-time metric in the exterior
region to the Schwarzschild radius $r_s$ is:
\[
ds^2 = b(r) d\bar{t}^2 - \frac{1}{b(r)} dr^2 \ , \qquad b(r)
\equiv 1 - \frac{r_s}{r}
\]
We can introduce the stationary time $t$ and the radial coordinate
$\rho$ from the horizon, both in units of Schwarzschild radius:
\begin{equation} \label{IV-t-rho}
t = \frac{\bar{t}}{r_s} \, , \qquad \quad
\rho = \frac{r}{r_s} -1 > 0
\end{equation}

To obtain a conformally flat expression for the metric,
one can change the radial coordinate $\rho$ for the new one $x$:
\begin{equation} \label{IV-x-rho}
x= x(\rho) \equiv \rho + \ln \rho
\end{equation}
The inverse function that gives the coordinate $\rho$ in terms of $x$
may be expressed by using the Lambert function $w = W(z)$, which
is defined as the inverse of the function $z = w \exp w$. Indeed,
from (\ref{IV-x-rho}) we have:
\begin{equation} \label{IV-rho-x}
\rho = \rho(x) \equiv W(\exp x)
\end{equation}
Finally, one can define the {\em stationary} null coordinates
$\{\texttt{u}, \texttt{v}\}$ [see Fig.
\ref{fig:schwarzschild-1}(a)]:
\begin{equation}\label{eq-uvtx}
\begin{array}{ll}
\texttt{u} = t + x       \  & \qquad \qquad  t = \frac12
(\texttt{u} + \texttt{v}) \
\\[1mm]
\texttt{v} = t - x   \ &  \qquad \qquad  x = \frac12 (\texttt{u} -
\texttt{v}) \
\end{array}
\end{equation}
With all these transformations one obtains:

{\em In the exterior region, the metric of Schwarzschild plane takes
the expression:
\begin{eqnarray} \label{IV-metric-uv}
 ds^2&=&r_s^2 \, b(\rho)\, d\texttt{{\rm u}} \, d\texttt{{\rm v}}
\\[3mm]
 b(\rho) \equiv \frac{\rho}{\rho + 1} \, , & \ & \rho \equiv
W\left(\exp\left[\frac{\texttt{{\rm u}} - \texttt{{\rm
v}}}{2}\right]\right)  \label{IV-b-rho-x}
\end{eqnarray}
where $W(z)$ is the Lambert function.}

The trajectories of two stationary emitters $\gamma_i$ are defined
by the conditions [see Fig. \ref{fig:schwarzschild-1}(a)]:
\begin{equation} \label{IV-rho-is}
\rho = \rho_i \ , \qquad 1 < \rho_2 < \rho_1
\end{equation}
The method exposed in Sec. \ref{section-previous} to introduce
emission coordinates is based in obtaining the metric tensor in a
null coordinate system, as expressions (\ref{IV-metric-uv}) and (\ref{IV-b-rho-x}) do, and in
knowing the proper time history of the emitters in this system. In
coordinates $\{\texttt{u}, \texttt{v}\}$ for the two trajectories
(\ref{IV-rho-is}) one has, by (\ref{IV-x-rho}) and
(\ref{eq-uvtx}), $\dot{\texttt{u}} = \dot{\texttt{v}}$ and then
from (\ref{IV-metric-uv}) the unit character of their velocity
gives $r_s^2 \, b(\rho_i)\,\dot{\texttt{u}}^2 = 1 $. Consequently,
the two trajectories are given by (the repeated index does not
indicate summation):
\begin{equation}  \label{Schprincipalemit}
\gamma_i \equiv \begin{cases} \, \texttt{u} = u_i(\tau^i) =
\lambda_i (\tau^i- \tau^i_0) + x_i
\\[1mm] \, \texttt{v} = v_i(\tau^i) = \lambda_i (\tau^i- \tau^i_0) -
x_i\end{cases}
\end{equation}
where the constants $\lambda_i$ and $x_i$ depend exclusively on The
Schwarzschild radius $r_s$ and the radial parameter $\rho_i$:
\begin{equation} \label{IV-lambda}
\lambda_i = \frac{1}{r_s}\sqrt{\frac{\rho_i + 1}{\rho_i}} \ ,
\qquad x_i = x(\rho_i)
\end{equation}
and where $\tau^i_0$ are the proper times that indicate the emitter
clocks at two events with the same Schwarzschild time $t$.

It is important to remark that {\em the radial parameter $\rho_i$
is a dynamical characteristic of the trajectory, equivalent to the
acceleration parameter $\alpha_i$}. Indeed, as the emitter
acceleration scalars are constant because emitter trajectories are
tangent to a Killing vector, a standard computation of them gives:
\begin{equation} \label{IV-accelerations}
\alpha_i =  \frac{1}{2 \, r_s \, \rho_i^{1/2} (\rho_i + 1)^{3/2}}
\end{equation}
Is is easy to see that this relation admits a unique real positive
solution $\rho_i = \rho_i(\alpha_i),$ showing their one-to-one
equivalence and, thus, the stated dynamical character of $\rho_i.$

\subsection{Emission coordinates and emitter trajectories}

According to (\ref{coordinatechange0}), the emission coordinates
$\{\tau^1,\tau^2\}$ are defined by the transformation to the
stationary null system $\{\texttt{u},\texttt{v}\}$:
\begin{equation}  \label{IV-cochange-a}
\begin{array}{l}
\texttt{u}  =  u_1(\tau^1) = \lambda_1 (\tau^1 - \tau^1_0) + x_1 \\[2mm]
\texttt{v} = v_2(\tau^2)  = \lambda_2 (\tau^2 - \tau^2_0) -  x_2
\end{array}
\end{equation}
and the inverse transformation is:
\begin{equation}  \label{IV-cochange-b}
\begin{array}{l}
\displaystyle \tau^1  = \tau^1_0 +  \frac{1}{\lambda_1}(\texttt{u}- x_1)\\[2mm]
\displaystyle \tau^2 =  \tau^2_0 + \frac{1}{\lambda_2}(\texttt{v}
+ x_2)
\end{array}
\end{equation}
\begin{figure}[htb]
    \includegraphics[angle=0,width=0.48\textwidth]{%
        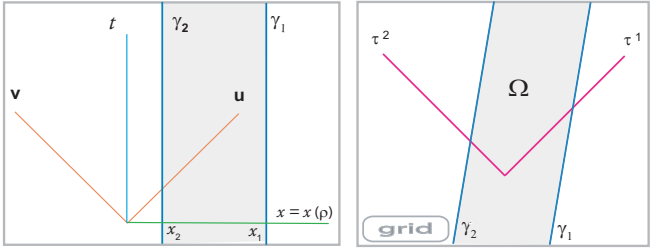}
    \caption{       \label{fig:schwarzschild-1}
        Trajectories of stationary emitters in Schwarzschild plane. (a) In
        `stationary' null coordinates $\{\texttt{u}, \texttt{v}\}$ associated with the static
        time $t$ and the coordinate $x=x(\rho)$. The stationary observers are defined by
        $\texttt{v} = \texttt{u} + constant$. (b) In the emission grid $\{\tau^1,\tau^2\}$.}
\end{figure}

Making use of (\ref{phis}), we can obtain the expression of the
emitter trajectories in emission coordinates $\{\tau^1,\tau^2\}$:
\begin{st} \label{IV-recti-trayec}
In Schwarzschild plane, the trajectories $\gamma_1$,
$\gamma_2$ of the emitters of a stationary positioning system in
emission coordinates $\{\tau^1,\tau^2\}$ are parallel straight
lines of the form:
\begin{eqnarray}  \label{IV-emitt-taus-a}
\gamma_1 \equiv \begin{cases} \tau^1 = \tau^1 \\ \tau^2 =
\frac{1}{\omega} (\tau^1 - q - \sigma) \equiv \varphi_1(\tau^1)
\end{cases}\,
\\[1mm]
\gamma_2 \equiv \begin{cases} \tau^1 = \omega \tau^2 - q + \sigma
\equiv \varphi_2(\tau^2)\\ \tau^2 = \tau^2\end{cases} \quad
\label{IV-emitt-taus-b}
\end{eqnarray}
The {\em slope parameter} $\omega$, the {\em separation parameter}
$q$ and the {\em synchronization parameter} $\sigma$ are related to
the Schwarzschild radius $r_s$, the radial parameters $\rho_i$ and
the synchronization instants $\tau^i_0$ of the emitters by:
\begin{equation} \label{IV-pi-q-sigma}
\begin{array}{l}
\displaystyle \omega  = \frac{\lambda_2}{\lambda_1} = \omega(\rho_1, \rho_2) \\[3.5mm]
\displaystyle q  = \frac{x_1 - x_2}{\lambda_1} = r_s Q(\rho_1,
\rho_2)\\[3.5mm]
\sigma = \tau^1_0 - \omega  \tau^2_0
\end{array}
\end{equation}
$\omega(\rho_1, \rho_2)$ and $Q(\rho_1, \rho_2)$ being the
bi-parametric expressions:
\begin{equation} \label{IV-pi-Q-rhos}
\begin{array}{l}
\displaystyle \omega(\rho_1, \rho_2) \equiv
\sqrt{\frac{\rho_1(\rho_2
+ 1)}{\rho_2(\rho_1 + 1)}} \\[4mm]
\displaystyle Q(\rho_1, \rho_2) \equiv \sqrt{\frac{\rho_1}{\rho_1
+ 1}} \left[\rho_1-\rho_2 + \ln \frac{\rho_1}{\rho_2}\right]
\end{array}
\end{equation}
\end{st}

Fig.  \ref{fig:schwarzschild-1}(b) illustrates these trajectories
in the grid.

\subsection{Emitter's positioning and dynamical parameters}

As a consequence of the above statement, any user traveling on the
emission coordinate domain $\Omega$ of this positioning system and
receiving the emitter positioning data $\{\tau^1, \tau^2;
\bar{\tau}^1, \bar{\tau}^2\}$ can find a particular linear relation
between the $\bar{\tau}$'s and the $\tau$'s, leading him to find
that the emitter trajectories $\varphi_i(\tau^i)$ are parallel
straight lines in the grid of the emission coordinates he is
receiving. From them, he can extract the parameters $\omega
> 1$, $q > 0$ and $\sigma$ [see (\ref{IV-emitt-taus-a}) and
(\ref{IV-emitt-taus-b})]:
\begin{st} \label{IV-two-events}
The slope, separation and synchronization parameters $\omega > 1$,
$q
> 0$ and $\sigma,$ characterizing the parallel trajectories in the
grid $\{\tau^1\}\times\{\tau^2\}$ of the stationary emitters in
Schwarzschild plane have, in terms of the emitter positioning data
at two different events $P$ and $Q$, the same expression as in
Minkowski plane, given by {\rm (\ref{III-par-taus})}.
\end{st}

Once these parameters are evaluated, expressions
(\ref{IV-pi-q-sigma}) and (\ref{IV-pi-Q-rhos}) tell us that $\omega$
and $q$ determine one-to-one the dynamical radial parameters
$\rho_i$, which characterize the emitter trajectories. Indeed, from
the first expression in (\ref{IV-pi-Q-rhos}) we can obtain:
\begin{equation} \label{IV-rho2}
\rho_2 = \frac{\rho_1}{(\omega^2-1) \rho_1 + \omega^2}
\end{equation}
and, substituting in the second one, we have:
\begin{equation} \label{IV-Q-rho1}
\displaystyle Q(\rho_1; \omega) \equiv \sqrt{\frac{\rho_1}{\rho_1 +
1}} \left[\frac{\rho_1(\rho_1 + 1)}{\rho_1 +
\frac{\omega^2}{\omega^2 - 1}} + \ln [(\omega^2-1) \rho_1 +
\omega^2]\right]
\end{equation}
This expression of $Q(\rho_1, \omega)$ is an effective function on
$\rho_1$. Consequently it admits an inverse and this radial
parameter may be obtained as a function of $\omega$ and $ Q \equiv
q/r_s$.

Moreover, the third expression in (\ref{IV-pi-q-sigma}) shows that the
parameter $\sigma$ gives the emitter synchronization. All these
results can be summarized in the following:

\begin{st} \label{IV-rhos-taus}
For stationary positioning systems in Schwarzschild plane, the
radial parameters $\rho_1$, $\rho_2$ of the emitters may be obtained
in terms of the slope data parameter $\omega > 1$ and the separation
data parameter $q > 0,$ by {\rm (\ref{IV-rho2})} and
\begin{equation} \label{IV-rhos-pi-q}
\rho_1 = \rho_1(\omega, q/r_s)\, ,
\end{equation}
where $\rho_1(\omega, q/r_s)$ is the inverse of {\rm
(\ref{IV-Q-rho1})}.

Also, their synchronization times $\tau^1_0$ and $\tau^2_0$
indicated by the emitter clocks at two simultaneous events, are
related to $\omega$ and the synchronization data parameter $\sigma$
by
\begin{equation} \label{IV-sigma}
\tau^1_0 = \sigma + \omega \tau^2_0
\end{equation}
\end{st}

This statement gives the parameters $\rho_1$ and $\rho_2$, as well
as the relation between $\tau^1_0$ between $\tau^2_0,$ in terms of
the data parameters $\omega,$ $q$ and $\sigma$, which can be
extracted from the data received by the user. In analogy to
Minkowskian case, these parameters yield an operational definition
of the null canonical stationary coordinates
$\{\texttt{u},\texttt{v}\}$ associated with the emitters and given
by (\ref{IV-cochange-a}).  By (\ref{IV-rhos-pi-q}) and
(\ref{IV-sigma}) these coordinates depend on a unique  parameter
(say $\tau^2_0$), which now controls the time-like translations in
the direction $\texttt{u} + \texttt{v}$.

\subsection{Metric in emission coordinates}

What is the metric information that the emitter positioning data
offer to the users? To obtain the metric function in emission
coordinates, starting  from (\ref{IV-metric-uv}) and using the
coordinate transformation (\ref{IV-cochange-a}), one obtains:
\begin{equation}  \label{IV-metric-0}
\begin{array}{lcl}
m(\tau^1, \tau^2) & = & r_s^2 \, b(\tau^1, \tau^2) u_1'(\tau^1)
v_2'(\tau^2) \\[2mm] & = & r_s^2\, b\left(\rho(x(\tau^1, \tau^2))\right) \lambda_1
\lambda_2
\end{array}
\end{equation}
But from (\ref{IV-cochange-a}) and (\ref{IV-pi-q-sigma}) we have:
\begin{equation}  \label{IV-metric-x}
\begin{array}{lcl}
2x(\tau^1, \tau^2) & = & u_1(\tau^1) - v_2(\tau^2) \\[2mm] & = &
 \lambda_1 (\tau^1 - \omega \tau^2 - \sigma - q) + 2x(\rho_1)
 \end{array}
\end{equation}
So that, by using (\ref{IV-lambda}) and (\ref{IV-rhos-pi-q}), we can
write the metric function in terms of the data $\{\tau^1, \tau^2\}$
and the parameters $\{\omega, q, \sigma \}$. Moreover these
parameters can be obtained as in (\ref{III-par-taus}) from the
emitter positioning data $\{\tau^1_P, \tau^2_P; \bar{\tau}^1_P,
\bar{\tau}^2_P\}$ and $\{\tau^1_Q, \tau^2_Q; \bar{\tau}^1_Q,
\bar{\tau}^2_Q\}$ at two events $P$ and $Q$. Thus, we can state:
\begin{st} \label{IV-prop-metric}
In Schwarzschild plane, the space-time metric in emission
coordinates of a stationary positioning system is given, in terms of
the emitter positioning data $\{\tau^1, \tau^2; \bar{\tau}^1,
\bar{\tau}^2\},$ by:
\begin{equation} \displaystyle
\label{IV-metric} \displaystyle d s^2 = \frac{1+\rho_1}{\rho_1}
\omega \, b(\tau^1, \tau^2) \, d \tau^1 d \tau^2
\end{equation}
where
\[
\displaystyle b(\tau^1, \tau^2) \equiv \frac{W(\exp x(\tau^1,
\tau^2))}{1+W(\exp x(\tau^1, \tau^2))} \, ,
\]
$W(z)$ being the Lambert function and
\[
\displaystyle x(\tau^1, \tau^2) \equiv
\frac{1}{2r_s}\sqrt{\frac{1+\rho_1}{\rho_1}}(\tau^1 - \omega \tau^2
- \sigma - q) + \rho_1 \ln \rho_1 ,
\]
where the parameters $\{\omega, q, \sigma \}$ are given by {\rm
(\ref{III-par-taus})} in terms of the values of the data at two
events, and the constant $\rho_1$ is defined in {\rm
(\ref{IV-rhos-pi-q})}.
\end{st}

This statement shows that the sole public quantities $\{\tau^1,
\tau^2; \bar{\tau}^1, \bar{\tau}^2\}$ received by any user allow him
to know the space-time metric everywhere. It is worth remarking that
the user must be informed that {\em the emitter positioning data
come from a stationary positioning system in presence of a mass of
the Schwarzschild radius $r_s$}. Remark also that this information
does not require the specific value of the emitter radial parameters
$\rho_i$, which may be determined by the user from the relations
(\ref{IV-rho2}) and (\ref{IV-rhos-pi-q}).

As statements \ref{III-recti-trayec} and \ref{IV-recti-trayec} show,
the sole emitter positioning data $\{\tau^1, \tau^2; \bar{\tau}^1,
\bar{\tau}^2\}$ give the same picture in the grid in Minkowski and
in Schwarzschild planes. Consequently, the additional quantitative
information of The Schwarzschild radius $r_s$ is a necessary
information in order to obtain the metric and the dynamical
quantities. Moreover, as we have already commented for Minkowski
plane, the user's knowledge of the accelerations of the clocks is
generically necessary because different dynamics (including non
uniformly accelerated ones) may produce the same emitter
trajectories {\em in the grid} \cite{cfm-c}. On the other hand, we
will show in next section that the data of one of the emitter's
accelerations allows the user to determine Schwarzschild mass.

\begin{figure}[htb]
    \includegraphics[angle=0,width=0.48\textwidth]{%
        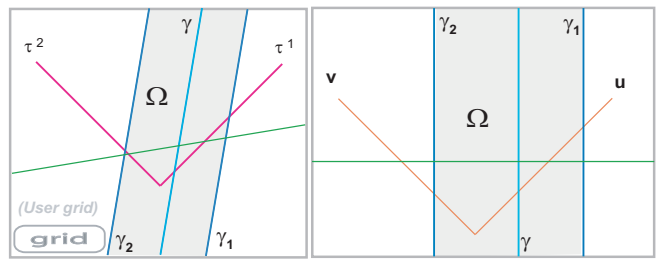}
    \caption{       \label{fig:schwarzschild-2}
        (a) User $\gamma$ with a trajectory in the grid parallel
        to the stationary emitters in Schwarzschild plane.
        Note that parallel straight lines $\tau^2 = -\frac{1}{\omega}
        \tau^1+ C$ are the hypersurfaces with Schwarzschild time
        $t=constant$. (b) Trajectory of the user $\gamma$ in `stationary' null coordinates
        $\{\texttt{u}, \texttt{v}\}$: it is an stationary user.}
\end{figure}

\subsection{Stationary user}

All the results above are independent of the user motion. Let us
consider now the specific user $\gamma$ which obtains from the
public data $\{\tau^1, \tau^2\}$ his own trajectory in the grid
as [see Fig. \ref{fig:schwarzschild-2}(a)]:
\begin{equation} \label{IV-user-a}
\tau^2 = F(\tau^1) = \frac{1}{\omega} (\tau^1 -c) \, , \qquad -q < c
- \sigma < q \, ,
\end{equation}
that is, a straight line parallel to the emitter trajectories. The
dynamics of this user and its relation with the emitters are given
by [see Fig. \ref{fig:schwarzschild-2}(b)]:

\begin{st} \label{IV-prop-user-estat-acc}
The users which find that their trajectory in the grid is a straight
line parallel to the emitter's ones {\rm [Eq. (\ref{III-user1-a})]}
are stationary users in Schwarzschild plane. Their radial parameter
$\bar{\rho}$ is:
\begin{equation} \label{IV-rho-bar}
\bar{\rho} = W(\exp \bar{x}), \qquad  \bar{x} \equiv
\frac12[x_1+x_2+\lambda_1(c-\sigma)] ,
\end{equation}
their constant acceleration scalar $\alpha$ is
\begin{equation} \label{Schaccelerationuser}
\alpha = \frac{1}{2 r_s \, \bar{\rho}^{1/2}(\bar{\rho} + 1)^{3/2}}
\end{equation}
and their proper time lapse $\Delta\tau$ is, for any of the two
emitter clocks (no summation)
\begin{equation} \label{IV-user-tau}
\Delta \tau =\frac{\lambda_i}{\lambda} \Delta \tau^i , \qquad
 \lambda = \frac{1}{r_s} \sqrt{\frac{\bar{\rho}+1}{\bar{\rho}}} .
\end{equation}
\end{st}

Observe that the parameters $x_i$, $\lambda_i$ and $\sigma$, and
then the value of the radial parameter $\bar{\rho}$ of the user
depend on the emitter positioning data $\{\tau^1, \tau^2;
\bar{\tau}^1, \bar{\tau}^2\}$ and The Schwarzschild radius $r_s$, as
a consequence of (\ref{IV-lambda}),  (\ref{IV-rho2}) and
(\ref{IV-rhos-pi-q}).

In order to show these relations, let us construct, starting
from the emitter positioning data, the stationary null coordinate
system $\{\texttt{u},\texttt{v}\}$ given by (\ref{IV-cochange-a}).
In it, using the inverse transformation (\ref{IV-cochange-b}), the
trajectory equation (\ref{IV-user-a}) becomes:
\begin{equation}  \label{IV-user-b}
\texttt{u} - \texttt{v} = 2 \bar{x}
\end{equation}
where $\bar{x}$ is given in (\ref{IV-rho-bar}). This means that this
trajectory belongs to the stationary family of trajectories as the
statement says.

Furthermore, the user can also obtain, up to an origin, his own
proper time $\tau$. Indeed, in the null coordinate system
$\{\texttt{u}, \texttt{v}\}$ the proper time user trajectory is:
\begin{equation}  \label{IV-user-traj-tau}
\gamma \equiv \begin{cases} \, \texttt{u} = \lambda (\tau -
\tau_0) + \bar{x}
\\ \, \texttt{v} = \lambda (\tau - \tau_0) - \bar{x}\end{cases}
\end{equation}
where the constant $\lambda$ is given by the second of expressions
(\ref{IV-user-tau}).

Then, (\ref{IV-user-traj-tau}) and the coordinate transformation
(\ref{IV-cochange-a})
 imply the first of expressions (\ref{IV-user-tau}).

\subsection{Simultaneous events for the emitters}

The simultaneity loci of the stationary emitters in stationary null coordinates are given by the lines $\texttt{u} + \texttt{v} = 2 t =
constant$. Using the coordinate transformation
(\ref{IV-cochange-a}) one obtains that these loci take in the grid the following expression:
\[
\tau^2 = - \frac{1}{\omega} \tau^1 + C .
\]
Thus, as in the flat case, one has:

\begin{st} \label{IV-prop-simultaneous}
The locus of simultaneous events for the emitters of a stationary
positioning system in Schwarzschild plane are parallel straight
lines of slope $-\frac{1}{\omega}$, the same, up to sign, as that of
the trajectory of the emitters.
\end{st}

\subsection{Comparing Minkowski}

It is worth remarking that the emitter positioning data received
by a user give, in the two physically different two-dimensional
Minkowski and Schwarzschild space-times, the same qualitative
information about the emitter trajectories in the grid, as we have
seen in this and the precedent section [see Figs.
\ref{fig:accelerated-1}(b) and \ref{fig:schwarzschild-1}(b)].
Moreover, the trajectory in the grid of any stationary user is
also the same in both cases [see Figs. \ref{fig:accelerated-2}(a)
and \ref{fig:schwarzschild-2}(a)]. In spite of this fact, we shall
see in the next section that complementary data that afford
dynamic information of the system allow the user to distinguish
between both systems.

The relevance of the dynamical data for the determination of the
system is manifest in the case of a geodesic user: his trajectory in
the grid is different in both cases. Is is easy to show this fact
because the trajectory (\ref{III-user-geo}) of a geodesic user in
the grid of the system in the flat case is not geodesic for the
Schwarzschild metric (\ref{IV-metric-uv}), as a straightforward
calculation shows. In this last case the geodesics, qualitatively
similar to those of the flat case shown in Fig.
\ref{fig:accelerated-3}(b), have nevertheless a different trajectory
equation. By comparison of this trajectory with the one constructed
by the geodesic user from the received sole data $\{\tau^1,
\tau^2\}$, he is able to distinguish in which space-time he is
evolving.

\section{Gravimetry and positioning: the information of the user data}
\label{section-gravimetry}

The interest of auto-locating positioning systems in gravimetry has
been pointed out in \cite{cfm-a}, where we have shown that, if a
user has neither a priori information on the gravitational field nor
on the positioning system, the user data determine the metric and
its first derivatives along the user and emitter trajectories. In
the next section we apply these results to obtain the gravitational
information that can be extracted from the user data generated by
the positioning systems of the precedent sections.

But in some cases the user can have some a priori information
about the space-time or about the kind of positioning system he is
using. The different levels of  not null information in these two
aspects establishes different kinds of gravity problems. For
example, in the last two precedent sections, we have supposed the
space-time fully known (Minkowski or Schwarzschild planes), and also the partial information on the positioning system assuring its
stationarity, but without specifying the space-time trajectories
of the emitters.

In this section we will consider other situations of total or
partial information on the space-time itself and/or on the
positioning system, and we will analyze the role that the emitter
accelerations and the proper user data can play in obtaining the
gravitational field and in recovering the complementary
characteristics of the positioning system.

We consider two levels in the a priori information that the user
can have about the gravitational field. He may know generically
that:
\begin{itemize}
\item[]
(GF-1) The space-time belongs to the family of Schwarzschild planes
expanded to include Minkowski plane,
\end{itemize}
or specifically that:
\begin{itemize}
\item[]
(GF-2) The space-time is a particular Schwarzschild plane with a
known the Schwarzschild radius $r_s$ or it is Minkowski plane.
\end{itemize}

We also consider two levels in the a priori user information about
the positioning system. He may know generically that:
\begin{itemize}
\item[]
(PS-1) The positioning system is defined by two stationary emitters,
\end{itemize}
or specifically that:
\begin{itemize}
\item[]
(PS-2) The positioning system is defined by two particular
stationary emitters with known radial coordinates $\rho_i$ and a
fixed synchronization $\sigma$.
\end{itemize}

\subsection{A user with full a priori information}

If the user has the widest information (GF-2) about the
gravitational field and the widest information (PS-2) about the
positioning system, then, from (GF-2) he can take a convenient
coordinate system and express in it the metric components and from
(PS-2) he can describe in this coordinate system the trajectory of
the emitters and the future light rays that spread from them. Then,
he may construct the emission coordinates and calculate the
coordinate transformation relating them to the initial ones.
Consequently, he can evaluate the emitter trajectories and the
space-time metric in emission coordinates. Thus, without receiving
any data, the user can know by calculation the pairs $\{\tau^1,
\bar{\tau}^2\}$ and $\{\tau^2, \bar{\tau}^1\}$ locating the emitters
as well as their dynamics.

With this full information about the metric and the positioning
system, a user receiving the user positioning data $\{\tau^1,
\tau^2\}$ may extract his trajectory $\tau^2 = F(\tau^1)$ in the
grid.  Moreover at every event $(\tau^1, \tau^2)$ on his
trajectory, he may know the metric function $m(\tau^1, \tau^2)$
and his acceleration.

Thus, in particular, the user may obtain his own units of time and
distance. For example, in the particular case of a stationary user,
his proper time is given by (\ref{IV-user-tau}).

\subsection{A user with  full metric information and partial knowledge
of the positioning system}

The situation in which a user has the widest information about the
gravitational field (GF-2) and partial information about the
positioning system (PS-1) is the one that has been analyzed in
detail in Sec. \ref{section-III} for Minkowski plane and in Sec.
\ref{section-IV} for Schwarzschild plane.

In particular, we have shown that the sole emitter positioning data
$\{\tau^1, \tau^2; \bar{\tau}^1, \bar{\tau}^2\}$ complete the system
information [see statements \ref{III-alfas-taus} and
\ref{IV-rhos-taus}]. Moreover we have shown that the space-time
metric may be obtained in terms of these data (see statements
\ref{III-prop-metric} and \ref{IV-prop-metric}).

Once completed the system information in this way, the user is in
the situation of the precedent subsection and, as commented there,
he can also obtain his trajectory and his own units of time and
distance.

\subsection{Obtaining The Schwarzschild mass from the emitter
accelerations}

Suppose now that the user has partial information (GF-1) about the
gravitational field and also partial information (PS-1) about the
positioning system. The user knows thus that the system is defined
by stationary emitters and, consequently, that they have constant
acceleration. If the emitters carry accelerometers and broadcast
their accelerations, the user can receive the precise value
$\alpha_1$ $\alpha_2$ of these accelerations.

Can this dynamical information determine The Schwarzschild mass? The
answer is affirmative. Furthermore, apart from the emitter
positioning data $\{\tau^1, \tau^2; \bar{\tau}^1, \bar{\tau}^2\}$,
{\em only one of the two accelerations} is sufficient to the user to
obtain The Schwarzschild mass and, consequently, to acquire the
information level considered in the precedent subsections.

In order to specify exactly this result let us assume firstly that
the user knows the full set of the public data  $\{\tau^1, \tau^2;
\bar{\tau}^1, \bar{\tau}^2; \alpha_1, \alpha_2\}$. We know that the
data parameters $\{\omega, q, \sigma\}$ depend on the emitter
positioning data [see (\ref{III-par-taus})]. Because the
accelerations $\alpha_i$ are also known, the first expression in
(\ref{IV-pi-Q-rhos}) and the the two in (\ref{IV-accelerations})
constitute three relations in $\{r_s, \rho_1, \rho_2\}$ that allow
to obtain them explicitly in terms of $\{\omega, \alpha_1,
\alpha_2\}$. On the other hand, the resulting expressions have to be
compatible with the expression of $q$ given in
(\ref{IV-pi-q-sigma}).

In the case of Minkowski plane, the user data parameters $\{\omega,
q, \alpha_1, \alpha_2\}$ are submitted not to one but to two
compatibility conditions, those given by (\ref{III-parametres-a}).
This fact and the considerations in the above paragraph lead to the
following

\begin{st} \label{prop-mass-accelerations}
Consider a user of a stationary positioning system in a space-time
of which he only knows that it is either a Schwarzschild or
Minkowski plane. Let $\{\tau^1, \tau^2; \bar{\tau}^1, \bar{\tau}^2;
\alpha_1, \alpha_2\}$ be the public data that he receives,
$\{\omega, q, \sigma\}$ the data parameters that can be extracted
from them by means of {\rm (\ref{III-par-taus})}, and $\kappa$ the
data parameter given by:
\begin{equation} \label{kappa-def}
\kappa \equiv \sqrt{\frac{ \omega \alpha_1}{\alpha_2}}.
\end{equation}
Then $\kappa$ is necessarily such that:
\begin{equation} \label{kappa}
\kappa  \leq 1.
\end{equation}
Moreover:\\

(i) The value $\kappa = 1$ is the one that informs the user that
his space-time is Minkowski plane. In this case, the public data
are submitted to the constraints:
\begin{equation} \label{restrictions-flat}
\alpha_1 \omega =  \alpha_2  \, , \qquad q \alpha_1 = \ln \omega
\end{equation}

(ii) Any value $\kappa < 1$ informs the user that his space-time
is Schwarzschild plane with radius
\[
r_s = \frac{(\omega^2 - 1)^2}{2 \, \alpha_2 (1-
\kappa)^{1/2}(\omega^2 - \kappa)^{3/2}},
\]
that the radial coordinate of the emitters are:
\[
\rho_1 = \frac{\omega^2 (1- \kappa)}{\kappa (\omega^2 - 1)} \, ,
\qquad \rho_2 = \frac{1- \kappa}{\omega^2 - 1},
\]
and that the public data are submitted to the constraint:
\begin{equation} \label{restriction}
2 q \alpha_1 \frac{\omega^2 - \kappa}{\kappa (\omega^2 - 1)} = 1 -
\kappa + \frac{ \kappa (\omega^2 - 1)}{\omega^2 - \kappa} \ln
\frac{\omega^2}{\kappa}
\end{equation}
\end{st}

The compatibility condition (\ref{restriction}) for the data
parameters $\{\omega, q, \alpha_1, \alpha_2\}$ suggests that the
emitter positioning data $\{\tau^1, \tau^2; \bar{\tau}^1,
\bar{\tau}^2\}$ and a sole acceleration, say $\alpha_1$, are
sufficient data to determine The Schwarzschild mass. Indeed, from
(\ref{IV-accelerations}) we obtain:
\begin{equation} \label{radiSch}
r_s =  \frac{1}{2 \, \alpha_1 \, \rho_1^{1/2} (\rho_1 + 1)^{3/2}}
\end{equation}
This expression allows eliminate $r_s$ in the expression
(\ref{IV-pi-q-sigma}) of $q$, which becomes:
\begin{eqnarray}
q&=&\frac{1}{2 \alpha_1} \Theta(\rho_1, \rho_2)  \label{q-alpha1} \\
\label{Theta}  \Theta(\rho_1, \rho_2)&\equiv&\frac{1}{(\rho_1 +
1)^2} \left[\rho_1-\rho_2 + \ln \frac{\rho_1}{\rho_2}\right]
\end{eqnarray}

Then, substituting (\ref{IV-rho2}) in (\ref{Theta}),
(\ref{q-alpha1}) takes the form:
\begin{equation}\label{rho1-implicit}
2 q \alpha_1 = \Theta(\rho_1, \omega) \, .
\end{equation}
This expression of $\Theta(\rho_1, \omega)$ is an effective function
on $\rho_1.$ Consequently it admits an inverse:
\begin{equation} \label{rho1-Rs}
\rho_1 = \rho_1 \left(\omega, q \alpha_1 \right) \, ,
\end{equation}
and we can state:

\begin{st} \label{prop-mass-acceleration}
Consider a user of a stationary positioning system in a space-time
of which he only knows that it is a Schwarzschild plane. Let
$\{\tau^1, \tau^2; \bar{\tau}^1, \bar{\tau}^2\}$ be the emitter
positioning data   that he receives and suppose that he also
receives but one of the emitter dynamical data, say $\alpha_1$.
Then, The Schwarzschild radius $r_s$ is given by {\rm
(\ref{radiSch})} where $\rho_1$ may be extracted from {\rm
(\ref{rho1-implicit})} as {\rm (\ref{rho1-Rs})}, in terms of the
data parameters $\{\omega,q, \alpha_1\}$.
\end{st}

\subsection{Obtaining The Schwarzschild mass from the user's
proper time}

Let us assume again that the user has partial information (GF-1) and
(PS-1) respectively about the gravitational field and the
positioning system. We show here that if the user is stationary and
generates his proper time, he can also obtain the Schwarzschild
mass.

Let us suppose that a user receiving the emitter positioning data
$\{\tau^1, \tau^2; \bar{\tau}^1, \bar{\tau}^2\}$ carries a clock
that measures his proper time $\tau$ and that he is stationary
with respect to the stationary positioning system. Remember that,
from statement \ref{III-prop-user-estat-acc}, he knows that he is
stationary if his trajectory in the grid is parallel to the
trajectories of the stationary emitters and that this property may
be extracted from the user positioning data he is receiving. From
his trajectory in the grid, he may also extract by
(\ref{IV-user-a}) the value of the specific data parameter $c$.

On the other hand, by comparing the user time $\tau$ with one of
the received times, say $\tau^1$, the user can obtain the data
parameter:
\begin{equation} \label{p-taus}
p = \frac{\Delta \tau^1}{\Delta \tau}
\end{equation}
This parameter relates the radial coordinates of the user and the
emitter $\gamma_1$. Indeed,  from (\ref{IV-lambda}) and
(\ref{IV-user-tau}) one obtains:
\begin{equation} \label{rouser-ro1}
\bar{\rho} = \frac{\rho_1}{(p^2-1) \rho_1 + p^2}
\end{equation}

Then, from the second expression in (\ref{IV-pi-q-sigma}) and
relations (\ref{IV-rho-bar}) and (\ref{IV-user-tau}) we have:
\begin{equation} \label{chi}
\frac{c-\sigma}{q} = \frac{2 \bar{x} - x_1 - x_2}{x_1 - x_2}
\equiv \chi(\rho_1, \rho_2, \bar{\rho}),
\end{equation}
and substituting (\ref{IV-rho2}) and (\ref{rouser-ro1}) one
obtains:
\begin{equation}
\frac{c-\sigma}{q} = \chi(\rho_1; \omega, p) \, .
\end{equation}
This expression of $\chi(\rho_1; \omega, p)$ is an effective
function on $\rho_1$. Consequently, it admits an inverse:
\begin{equation} \label{rho1-pi-p-c}
\rho_1 = \rho_1(\omega, p, \frac{c-\sigma}{q}) \, .
\end{equation}
Thus, we can state:

\begin{st} \label{prop-mass-time}
In a stationary positioning system of a space-time of which it is
only known that it is a Schwarzschild plane, let us consider a
stationary user receiving the emitter positioning data
$\{\tau^1,\tau^2; \bar{\tau}^1, \bar{\tau}^2\}$ and carrying a
clock measuring his proper time $\tau$. Then, the
Schwarzschild radius is:
$$
r_s = \frac{q}{Q(\rho_1;\omega)} ,
$$
where $\rho_1$ and then $Q(\rho_1;\omega)$ are given in terms of the
data parameters $\{\omega, q, \sigma, c\}$ respectively by {\rm
(\ref{rho1-pi-p-c})} and {\rm (\ref{IV-Q-rho1})}.
\end{st}

\section{The gravimetry general problem: obtaining the metric
along trajectories} \label{section-grav-along}

Now we suppose that, in a Schwarzschild or Minkowski space-time,
and in a stationary positioning system, we have a user which is
unaware of these two facts, i.e. he has no information about the
gravitational field and the positioning system.

We know that the emitter positioning data $\{\tau^1, \tau^2;
\bar{\tau}^1, \bar{\tau}^2\}$ determine the user trajectory
$\tau^2 = F(\tau^1)$ as well as the emitter trajectories
$\bar{\tau}^j = \varphi_i(\tau^i)$ in the grid, and that this
happens whatever be the curvature of the two-dimensional
space-time.

In \cite{cfm-a} we have studied the information that these data
give about the space-time metric and we have shown that {\em the
emitter positioning data $\{\tau^1, \tau^2; \bar{\tau}^1,
\bar{\tau}^2\}$ determine the space-time metric function along the
emitter trajectories}. If in addition the user knows the emitter
dynamical data $\{\alpha_1, \alpha_2\}$, and consequently the
acceleration scalars $\alpha_1(\tau^1)$, $\alpha_2(\tau^2)$, we
have also shown in \cite{cfm-a} that {\em the public data
$\{\tau^1, \tau^2; \bar{\tau}^1, \bar{\tau}^2; \alpha_1,
\alpha_2\}$ determine the gradient of the space-time metric
function along the emitter trajectories}.

We can apply the general explicit expressions given in \cite{cfm-a}
and the results of the sections \ref{section-III} and
\ref{section-IV} to obtain the metric component and its gradient
corresponding to the parallel straight lines emitter trajectories
and constant accelerations broadcast by the stationary emitters. The
result is:

\begin{st} \label{prop-along-emitters}
In a Schwarzschild or Minkowski space-time, and in a stationary
positioning system, a user unaware of these two facts can obtain:\\

(i) From the emitter positioning data  $\{\tau^1, \tau^2;
\bar{\tau}^1, \bar{\tau}^2\}$, the metric function on the emitter
trajectories:
\begin{equation} \label{metric-along-emitters}
m(\tau^1) = \omega \ , \quad  m(\tau^2) = \frac{1}{\omega}
\end{equation}

(ii) From the public data $\{\tau^1, \tau^2; \bar{\tau}^1,
\bar{\tau}^2,\alpha_1, \alpha_2\}$, the gradient of the metric
function along the emitter trajectories:
\begin{equation}  \label{gradmetric-along-emitters}
\begin{array}{l}
(\ln m),_1 (\tau^1) =\alpha_1 \ ,  \quad (\ln m),_2 (\tau^1) = -
\omega
\alpha_1\\[2mm]
(\ln m),_1 (\tau^2) = \omega \alpha_2 \ ,  \quad (\ln m),_2 (\tau^2)
= - \alpha_2
\end{array}
\end{equation}
where $\omega$ is obtained from the emitter positioning data by {\rm
(\ref{III-par-taus})}.
\end{st}

It is worth remarking that the same public data giving the metric
(\ref{metric-along-emitters}) and its gradient
(\ref{gradmetric-along-emitters}) along the stationary emitter
trajectories can be generated by others, not necessarily stationary,
positioning systems  and/or in  space-times different from
Schwarzschild planes. It follows that the compatibility restrictions
(\ref{kappa}) and (\ref{restriction}) [or (\ref{restrictions-flat})
for Minkowski plane] are necessary but not sufficient conditions to
characterize a stationary positioning system in Schwarzschild
planes. Consequently, the user needs to receive complementary data
or a priori information in order to obtain the gravitational field
everywhere or to get the specific characteristics of the positioning
system.

Alternatively, even if the positioning system is not auto-locating
(broadcasting the sole proper times $\{\tau^1, \tau^2\}$), when the
user also knows the proper user data $\{\tau, \alpha\}$, he may
obtain gravimetric information along his trajectory. Indeed, from
these data the user can obtain his parameterized proper time
trajectory (\ref{user-proper-time}) and his acceleration scalar
$\alpha(\tau)$. We have then shown in \cite{cfm-a} that {\em the
public-user data $\{\tau^1, \tau^2; \tau, \alpha\}$ determine the
space-time metric function and its gradient on the user trajectory}.

Thus, applying the explicit expressions given in \cite{cfm-a} to the
stationary user considered in the sections \ref{section-III} and
\ref{section-IV}, one has:

\begin{st} \label{prop-along-user}
In a Schwarzschild or Minkowski space-time, and in a stationary
positioning system, a user unaware of these two facts can obtain,
from the user positioning data $\{\tau^1, \tau^2\}$ and the proper
user data $\{ \tau, \alpha\}$, the metric function and its gradient
along the user trajectory:
\begin{eqnarray}  \label{metric-along-user}
m(\tau) = \frac{(\Delta \tau)^2}{\Delta \tau^1 \Delta \tau^2}
\qquad \qquad \qquad
\\[2mm]
(\ln m),_1 (\tau) = \alpha \frac{\Delta \tau}{\Delta \tau^1}  \ ,
\quad (\ln m),_2 (\tau) = - \alpha \frac{\Delta \tau}{\Delta
\tau^2} \label{gradmetric-along-user}
\end{eqnarray}
\end{st}

The user data that give the metric (\ref{metric-along-user}) and
the metric gradient (\ref{gradmetric-along-user}) along the user
trajectory can be received by a non stationary user of other
positioning systems and/or in other different space-times. Again,
the user needs complementary information in order to obtain the
metric everywhere and the characteristics of the positioning
system.

\section{Discussion and work in progress \label{discussion}}

In this work we have continued the two-dimensional approach to
relativistic positioning systems initiated in \cite{cfm-a}. In
that work we explained the basic features of these systems, we
studied in detail the positioning system defined by two geodesic
emitters in flat space-time, and we showed that in an arbitrary
space-time the user data determine the gravitational field and its
gradient along the emitters and user world lines.

Here we have considered stationary positioning systems, i.e.
positioning systems defined by two uniformly accelerated emitters at
rest with respect to each other. In order to understand the behavior
of the emission positioning systems in a non necessarily flat
gravitational field, we have studied stationary positioning systems
in  both, Minkowski (Sec. \ref{section-III}) and Schwarzschild
planes (Sec. \ref{section-IV}).

We have shown that a user that receives the emitter positioning data
will find, for the emitter trajectories, parallel straight lines in
the grid, no matter the plane be  (statements \ref{III-recti-trayec}
and \ref{IV-recti-trayec}),  and we have given the parameters of
these straight lines in terms of the emitter positioning data
(statements \ref{III-two-events} and \ref{IV-two-events}). We have
also obtained the dynamics (including the radial coordinate in the
Schwarzschild case) and synchronization of the emitters in terms of
data parameters (statements \ref{III-alfas-taus} and
\ref{IV-rhos-taus}). We have found that, if a user knows the
stationary character of the positioning system, the emitter
positioning data make him able to know explicitly the space-time
metric in emission coordinates (statements \ref{III-prop-metric} and
\ref{IV-prop-metric}). We have studied the proper data received by a
stationary user and the expressions of his acceleration and proper
time in terms of the emitter positioning data that he receives
(statements \ref{III-prop-user-estat-acc} and
\ref{IV-prop-user-estat-acc}). We have also obtained in the emission
grid the locus of simultaneous events for the emitters (statements
\ref{III-prop-simultaneous} and \ref{IV-prop-simultaneous}). In
Minkowski plane, the trajectory of a geodesic user in emission
coordinates has  been obtained (statement \ref{III-prop-inertial}).

The study of these two scenarios brings to light an interesting
situation: the emitter positioning data of both systems lead to an
identical emission grid. How a user can distinguish both systems? We
have analyzed this question in Sec. \ref{section-gravimetry} and
have shown that the knowledge of complementary user data determines
The Schwarzschild mass (statements \ref{prop-mass-accelerations},
\ref{prop-mass-acceleration} and \ref{prop-mass-time}). These simple
two-dimensional situations suggests that the relativistic
positioning systems could be useful in four-dimensional gravimetry
for reasonable parameterized models of the gravitational field.

The gravimetry cases analyzed here are only particular situations of
the general gravimetry problem in Relativity where user data are the
unique information that a user has. We showed in \cite{cfm-a} that
these data allow the user to obtain the metric function and its
gradient on emitter and on user trajectories. In Sec.
\ref{section-grav-along} we have applied these results to obtain the
gravitational information acquired by the user considered  in
sections \ref{section-III} and \ref{section-IV} (statements
\ref{prop-along-emitters} and \ref{prop-along-user}).

Finally, for a user that knows the space-time where he is immersed
(flat, Schwarzschild,...) but he has no information about
positioning system, we can ask the following questions. Can the user
data determine the characteristics of the positioning system? Can he
obtain information on his local units of time and distance and his
acceleration? The answer to these questions is still an open problem
for a generic space-time. But in a future paper \cite{cfm-c} we
undertake this query for Minkowski plane and we analyze the minimum
set of data that determine all the user and system information. A
remarkable result is that the user data are not independent
quantities: the accelerations of the emitters and of the user along
their trajectories are determined by the sole knowledge of the
emitter positioning data and of the acceleration of only one of the
emitters and only during the interval between the emission time of a
signal by an emitter and its reception time after being reflected by
the other emitter.

\ \\

\begin{acknowledgments}
This work has been supported by the Spanish Ministerio de
Educaci\'on y Ciencia, MEC-FEDER projects AYA2003-08739-C02-02 and
FIF2006-06062.
\end{acknowledgments}

\end{document}